\documentstyle[12pt,aaspp4]{article}

\begin{document}


\title{OBSERVATION OF MULTI-TeV DIFFUSE GAMMA RAYS FROM THE GALACTIC PLANE
WITH THE TIBET AIR SHOWER ARRAY}

\author{ M.~Amenomori\altaffilmark{1},  S.~Ayabe\altaffilmark{2},
        S.H.~Cui\altaffilmark{3},       L.K.~Ding\altaffilmark{3},
        X.H.~Ding\altaffilmark{4},      C.F.~Feng\altaffilmark{5}
        C.Y.~Feng\altaffilmark{6},      Y.~Fu\altaffilmark{5},
        X.Y.~Gao\altaffilmark{7},       Q.X.~Geng\altaffilmark{7},
        H.W.~Guo\altaffilmark{4},       M.~He\altaffilmark{5},
        K.~Hibino\altaffilmark{8},      N.~Hotta\altaffilmark{9},
        J.~Huang\altaffilmark{9},       Q.~Huang\altaffilmark{6},
        A.X.~Huo\altaffilmark{3},       K.~Izu\altaffilmark{10},
        H.Y.~Jia\altaffilmark{6},       F.~Kajino\altaffilmark{11},
        K.~Kasahara\altaffilmark{12},   Y.~Katayose\altaffilmark{13},
        K.~Kawata\altaffilmark{11},     Labaciren\altaffilmark{4},
        G.M.~Le\altaffilmark{14},       J.Y.~Li\altaffilmark{5},
        H.~Lu\altaffilmark{3},          S.L.~Lu\altaffilmark{3},
        G.X.~Luo\altaffilmark{3},       X.R.~Meng\altaffilmark{4},
        K.~Mizutani\altaffilmark{2},    J.~Mu\altaffilmark{7},
        H.~Nanjo\altaffilmark{1},       M.~Nishizawa\altaffilmark{15},
        M.~Ohnishi\altaffilmark{10},    I.~Ohta\altaffilmark{9},
        H.~Ooura\altaffilmark{11},      T.~Ouchi\altaffilmark{10},
        S.~Ozawa\altaffilmark{9},       J.R.~Ren\altaffilmark{3},
        T.~Saito\altaffilmark{16},      M.~Sakata\altaffilmark{11},
        T.~Sasaki\altaffilmark{8},      M.~Shibata\altaffilmark{13},
        A.~Shiomi\altaffilmark{10},     T.~Shirai\altaffilmark{8},
        H.~Sugimoto\altaffilmark{17},   K.~Taira\altaffilmark{17},
        M.~Takita\altaffilmark{10},     Y.H.~Tan\altaffilmark{3},
        N.~Tateyama\altaffilmark{8},    S.~Torii\altaffilmark{8},
        H.~Tsuchiya\altaffilmark{10},   S.~Udo\altaffilmark{2},
         T.~Utsugi\altaffilmark{2},      C.R.~Wang\altaffilmark{5},
        H.~Wang\altaffilmark{3},        X.~Wang\altaffilmark{5},
        X.W.~Xu\altaffilmark{3},        L.~Xue\altaffilmark{5},
        X.C.~Yang\altaffilmark{7},      Y.~Yamamoto\altaffilmark{11},
        Z.H.~Ye\altaffilmark{14},       G.C.~Yu\altaffilmark{6},
        A.F.~Yuan\altaffilmark{4},      T.~Yuda\altaffilmark{18},
        H.M.~Zhang\altaffilmark{3},     J.L.~Zhang\altaffilmark{3},
        N.J.~Zhang\altaffilmark{5},     X.Y.~Zhang\altaffilmark{5},
        Zhaxiciren\altaffilmark{4} and  Zhaxisangzhu\altaffilmark{4}\\
(The Tibet AS${\bf \gamma}$ Collaboration)}

\altaffiltext{1}{  Department of Physics, Hirosaki University, Hirosaki
036-8561, Japan}
\altaffiltext{2}{  Department of Physics, Saitama University, Saitama
338-8570, Japan}
\altaffiltext{3}{  Laboratory of Cosmic Ray and High Energy Astrophysics,
Institute of High Energy
Physics, Chinese Academy of Sciences, Beijing 100039, China}
\altaffiltext{4}{  Department of Mathematics and Physics, Tibet University,
Lhasa 850000, China}
\altaffiltext{5}{  Department of Physics, Shangdong University, Jinan
250100, China}
\altaffiltext{6}{  Institute of Modern Physics, South West Jiaotong
University, Chengdu 610031, Chi
na}
\altaffiltext{7}{  Department of Physics, Yunnan University, Kunming 650091,
China}
\altaffiltext{8}{  Faculty of Engineering, Kanagawa University, Yokohama
221-8686, Japan}
\altaffiltext{9}{  Faculty of Education, Utsunomiya University, Utsunomiya
321-8505, Japan}
\altaffiltext{10}{ Institute for Cosmic Ray Research, University of Tokyo,
Kashiwa 277-8582, Japan}
\altaffiltext{11}{ Department of Physics, Konan University, Kobe 658-8501,
Japan}
\altaffiltext{12}{ Faculty of Systems Engineering, Shibaura Institute of
Technology, Saitama 330-8570, Japan}
\altaffiltext{13}{ Faculty of Engineering, Yokohama National University,
Yokohama 240-0067, Japan}
\altaffiltext{14}{ Center of Space Science and Application Research, Chinese
Academy of Sciences, B
eijing 100080, China}
\altaffiltext{15}{ National Institute for Informatics, Tokyo 112-8640,
Japan}
\altaffiltext{16}{ Tokyo Metropolitan College of Aeronautical Engineering,
Tokyo 116-0003, Japan}
\altaffiltext{17}{ Shonan Institute of Technology, Fujisawa 251-8511, Japan}
\altaffiltext{18}{ Solar-Terrestrial Environment Laboratory, Nagoya
University, Nagoya 464-8601, Japan}

\begin{abstract}

Data from the Tibet-III air shower array (with energies around 3 TeV) and from
the Tibet-II array (with energies around 10 TeV) have been searched for diffuse
gamma rays from the Galactic plane.  These arrays have an angular resolution
of about 0.9 degrees.  The sky regions searched are the inner Galaxy,
$20^\circ \leq l \leq 55^\circ$, and outer Galaxy, $140^\circ \leq l \leq 225^\circ$,
and $|b| \leq 2^\circ$ or $\leq 5^\circ$. 
No significant Galactic plane gamma-ray excess was observed.
The 99\% confidence level upper limits for
gamma-ray intensity obtained are (for $|b| \leq 2^\circ$) 1.1 $\times
10^{-15}$ cm$^{-2}$s$^{-1}$sr$^{-1}$MeV$^{-1}$ at 3 TeV and 4.1 $\times
10^{-17}$ cm$^{-2}$s$^{-1}$sr$^{-1}$MeV$^{-1}$ at 10 TeV for the inner
Galaxy, and 3.6 $\times 10^{-16}$ cm$^{-2}$s$^{-1}$sr$^{-1}$MeV$^{-1}$ at
3 TeV and 1.3 $\times 10^{-17}$ cm$^{-2}$s$^{-1}$sr$^{-1}$MeV$^{-1}$ at 10 TeV
for the outer Galaxy, assuming a differential spectral index of 2.4.  The upper
limits are significant in the multi-TeV region when compared to
those from Cherenkov telescopes in the lower energy region and other air
shower arrays in the higher energy region; however, the results are not
sufficient to rule out the inverse Compton model with a source electron
spectral index of 2.0.

\end{abstract}


\section{INTRODUCTION}

  Detection of diffuse gamma rays from the Galactic plane is
considered to be a promising way to understand spatial distributions of
cosmic-ray acceleration regions, of interstellar matter (ISM), and
interstellar photon field (ISPF) densities.  Experimental data in the energy
region below 10 GeV with the EGRET instrument \cite{hunt97}, are fairly well
interpreted in terms of interactions of cosmic-ray hadrons and electrons
with ISM and ISPF.
However, a portion of the contribution of each elementary process of
interaction is uncertain due to assumed intensities and spectral indices of
cosmic-ray hadrons, electrons, and densities of ISM and ISPF.
With experiments in the energy region above 100 GeV, only upper limits for
diffuse gamma-ray intensities have been obtained, and these results
provide some constraints on the parameters in the models.  In order to
constrain the parameter values more severely in this
energy region, it is necessary to obtain more observational data.

\par
In the early stage of cosmic-ray astrophysics, the production rate of gamma
rays by Compton scattering of starlight photons by cosmic-ray electrons
was calculated by Feenberg \& Primakoff (1948), and gamma rays arising from
neutral pions through cosmic-ray collision with ISM was studied by Hayakawa
(1952).  Morrison (1958) advocated gamma-ray astronomy because of the
known production mechanisms and the straight trajectories from the origin.
Ginzburg \& Syrovatskii (1964) summarized the elementary process for
gamma-ray production and discussed diffusion and propagation of cosmic rays
in the Galactic disc.  As a result, it is quite natural to expect gamma rays
from the Galactic plane, however was twenty years after the
pioneering work by Feenberg \& Primakoff (1948) until the
intense band of gamma-ray intensity along the Galactic plane was observed
with OSO~3 \cite{krau72} and a balloon-borne detector \cite{fich72}.
Advanced observations were carried out
by detectors borne on SAS~2 (Fichtel et al. 1975; Hartman et al. 1979) and
COS~B (Mayer-Hasselwander et al. 1980, 1982).  These observations revealed
that the intensity profile of high-energy gamma rays is quite
relevant to the structure of the Galaxy, and have stimulated
much theoretical work in gamma-ray astronomy below 10 GeV (see, Bertsch et al.
1993 and references therein).

\par
A detailed intensity distribution of high-energy gamma rays coming from the
Galactic plane was given by Hunter et al. (1997) based on EGRET observations.
In the energy region above 1 GeV, the gamma-ray intensity from the inner
Galaxy is higher than the COS~B data by a factor of about 3.  It is also
higher than the conventional model predictions (e.g., Hunter et al. 1997;
Bertsch et al. 1993) by a factor of 1.7, where the neutral pion production
is based on the calculations of Stecker (1988), assuming the power law proton
spectrum with spectral index of 2.75 \cite{derm86}.   Mori (1997) showed
that the latter discrepancy can not be attributed to a calculation with
inaccurate accelerator data on neutral pion production. It can,
however, be interpreted by adopting a harder proton spectral index of 2.45
for the EGRET excess within a plane thickness of $|b| \leq 10^\circ$. Webber
(1999) also showed that the EGRET excess in $|b| \leq 5^\circ$ can be reproduced
by assuming a source proton spectral index of 2.25.

\par
 Following an indication by Schlickeiser (1979) of the importance of using
the Klein-Nishina cross section, Protheroe \& Wolfendale (1980) showed possible
dominance of inverse Compton (IC) gamma rays over
neutral-pion decay gamma rays for a range of electron injection spectra.
Detailed calculations of IC gamma rays above 50 MeV were made by Chi et al.
(1989), which were more than 50 \% of the total diffuse intensity at a
medium galactic latitude of $|b|=10^\circ - 20^\circ$.  Porter \& Protheroe
(1997) indicated that in such a high-energy region, cosmic-ray
electrons may create a significant part of the diffuse gamma rays,
depending on their injection spectral index and acceleration cutoff energy.
Pohl \& Esposito (1998) remarked that most of radio synchrotron spectra of
SNRs are well represented by power law indices around 0.5,
corresponding to an electron injection index of about 2.0 \cite{gree95}.
They argued that if this injection electron
index is employed, the EGRET excess above 1 GeV can be well explained by IC
scattering.  They also argued that an electron injection index of 2.0 is
within the expected Poisson fluctuations and is reasonable in the direction
toward the Galactic center with the line of sight passing through the
vicinity of many SNRs, taking into account the diffusion coefficient and the
observed local electron index of around 3.0 below 1 TeV.   Recently,
the energy spectrum of local electrons has been obtained with balloon-borne
instruments HEAT \cite{barw98} and BETS \cite{tori01}.

\par
Some models of diffuse Galactic gamma-ray continuum radiation below 10 GeV
were synthetically discussed by Strong, Moskalenko, \& Reimer (2000) with
respect to the injection spectra, including a harder nucleon and electron
spectral indices, and also with respect
to different ISM and ISPF densities.  At the higher energies of
the TeV-PeV region, the diffuse gamma-ray emission was calculated by
Berezinsky et al. (1993) in terms of cosmic-ray interaction with ISM, and
also by Ingelman \& Thunman (1996) using the current models for high energy
particle interaction. Broad-band diffuse gamma-ray emission, covering the
MeV-PeV region, was comprehensively calculated by Aharonian \& Atoyan
(2001), and they discussed propagation of hadron and electron
components and their injection spectra. Assuming that the hadron spectral
index is 2.15 (or 2.0) in SNRs, Berezhko \& V$\ddot{\rm o}$lk (2000)
showed that the averaged contribution to the diffuse gamma-ray flux should
exceed the current model predictions (Hunter et al. 1997; Bertsch et al 1993)
by a factor 5 (or 29) at 1 TeV.

\par
 Recently, several groups gave upper limits for diffuse gamma-ray fluxes
(Borione et al. 1998) above 100 TeV from the outer Galactic plane, using large
ground-based air shower arrays and muon detectors, and above 500 GeV
\cite{lebo00} and around 1 TeV \cite{aha01} from the inner Galactic plane, using
imaging atmospheric Cherenkov telescopes (IACT).  Using a small
scintillation counter array, Tibet-I, upper limits \cite{amen97} were given
at 10 TeV from the inner and outer Galactic planes.  In this paper we report
new upper limits from both the enlarged Tibet-II array and the high detector
density array of Tibet-III.

\section{EXPERIMENT}

After three years of operation, starting in 1990, of the Tibet-I array (located at
an altitude of 4300 m a.s.l. (606 gcm$^{-2}$) at Yangbajing
(90.53$^\circ$E, 30.11$^\circ$N) in Tibet of China), many scintillation
counters were added to this array late in 1994 to improve the sensitivity to allow
detection of $\sim$10 TeV gamma rays from a known source like the Crab Nebula,
and from other possible sources as well. The Tibet-II array
consists of 221 scintillation counters of 0.5 m$^2$ each, keeping the
same lattice interval of 15 m as Tibet-I.  The performance of the
Tibet-II array is almost same as the Tibet-I which is described elsewhere
(Amenomori et al. 1992, 1993).  The mode energy of all triggered air
showers is estimated from simulations to be about 7 TeV and 8 TeV for proton
initiated showers with a spectral index of 2.7 and gamma-ray initiated ones
assuming a spectral index of 2.5, respectively. The mode energy is 10 TeV
for air showers with $\Sigma \rho_{\rm FT} \geq$ 15 m$^{-2}$,
where $\Sigma \rho_{\rm FT}$/2 is the sum of the number
of particles that hit all 0.5 m$^2$ detectors.  This mode energy can
be calibrated by the magnitude of the westward shift of the moon shadow, and
the maximum deficit position due to the geomagnetic effect (Amenomori et al.
1993, 1996).  The angular resolution is determined to be 0.9$^\circ$ at 10
TeV, also using the deficit shape (depth and width) of the moon shadow.

\placefigure{fig1}

  The Tibet-II array covers 36,900 m$^2$, which is 4.5 times larger than
Tibet-I, except for some scattered detectors surrounding the lattice area in
each array \cite{amen95} as shown in Fig. 1.  The data acquisition rate is
about 200 Hz with the trigger condition requiring any four detectors to have 0.80
particles per detector. The observed number of air showers is 5.44 $\times
10^9$ events after the pre-analysis determining the arrival direction and
shower axis location, for the effective 551.2 days during the period 1997
February through 1999 September.  The event rate is about 114 events per
second in the Tibet-II data.

 In 1999, the array was enlarged, with many more scintillation counters so as
to make a high detector density array, Tibet-III, with a 7.5 m lattice
interval.  Fig. 1 shows the status in 2001.  The Tibet-III array
consists of 533 scintillation counters covering 22,050 m$^2$.  The full
Tibet-III array will be completed in 2002; it will cover 36,900 m$^2$, the same
as Tibet-II, as described in Amenomori et al. (2001).  The mode energy
of Tibet-III is about 3 TeV for proton initiated showers and the angular
resolution is 0.87$^\circ$ in the energy region above 3 TeV.  The data
acquisition rate is 680 Hz with the same trigger condition as described above
for Tibet-II.  The number of air showers is 1.35 $\times
10^{10}$ events, of which the arrival direction and shower axis location are
determined for the effective 517.3 days during the period 1999 November
through 2001 October.  The event rate is about 302 events per second in the
Tibet-III data. The performance of these arrays and the data obtained are
summarized in Table 1.


\begin{center}
TABLE 1\\
ARRAY PERFORMANCE AND OBTAINED DATA
\end{center}
\begin{tabular}{c|c c c c c c c c}\hline \hline
Air    & Lattice & Inner & Trigg & Observd & \multicolumn{3}{c}{Pre-analysis
for $\Sigma \rho_{\rm FT}\geq 15$}& Inner \\
Shower&Interval&Area & Rate$^\dagger$ & Period & Events & $E_{\rm mode}$ &
Ang. res. & Events \\
Array &(m)&(m$^2$) &(Hz)&Net days &(10$^9$)&(TeV)&(deg.) &(10$^9$)\\ \hline
       &         &           &      & 1997 Feb & & & & \\
Tibet-II&15 & 28,350 & 200 &\hspace*{.5mm}$\sim$99 Sep &5.44 &10 &0.9 & 4.14
\\
       &         &           &     & 551.2 & &&& \\ \hline
       &         &           &     & 1999 Nov & &  &   & \\
Tibet-III& 7.5 &22,050$^*$   & 680 &\hspace*{.5mm}$\sim$01 Oct &13.5 &3
&0.87 &6.59\\
      &         &            &     & 517.3 & & & & \\ \hline
\end{tabular}

{\small

\hspace*{-3mm}$^\dagger$ Trigger level is any four detectors with 0.80
particles per detector.\\[-8mm]

\hspace*{-3mm}$^*$ At the status in 1999$\sim$2002.  The inner area is
increased to $\sim$32,500 m$^2$ till the end of 2002.
}
\vspace*{5mm}

\par
Figure 2 shows an exposure map in galactic coordinates for air showers
obtained with the Tibet-III array, with zenith angles $\theta \leq
50^\circ$.  The shower event density increases from light gray to dark gray.
The Tibet-II array data produce a quite similar map. For the on-plane
data, shower events are employed in the sky regions of $20^\circ \leq l
\leq 55^\circ$ for the inner Galaxy (IG) and of $140^\circ \leq l \leq
225^\circ$ for the outer Galaxy (OG), in $|b| \leq 2^\circ$ or $\leq
5^\circ$ along the Galactic plane.

\placefigure{fig2}

\section{DATA ANALYSIS }

For the analysis in the energy range around 10 TeV, 4.14 $\times 10^9$ air
shower events are used with zenith angles $\theta \leq 50^\circ$, whose axes
hit the inner area of 28,350 m$^2$ of the Tibet-II array to assure the
quality of the arrival angle determination.
Around 3 TeV, 6.59 $\times 10^9$ events are used with $\theta \leq 50^\circ$
whose axes hit the 7.5 m lattice area of 22,050 m$^2$ of the Tibet-III
array.  The number of events that hit these inner areas are tabulated also
in Table 1 in the last column.  Those air shower events are assigned to the sky
regions from which they arrived, of 4$^\circ (10^\circ)$ bin 90 (36) belts
along the Galactic plane in equatorial coordinates both for the Tibet-II
and Tibet-III array data.  Figure 3 shows 10$^\circ$ bin warped belts for IG
with the declination range of $-10^\circ \leq \delta \leq 20^\circ$ and OG
with $-10^\circ \leq \delta \leq 60^\circ$. These declination ranges of the
on-plane belts
in the equatorial coordinates correspond to parts of a convex lens
shape, with a maximum thickness of about 4$^\circ$ or 10$^\circ$ in
galactic coordinates, as shown in Fig. 2.  The primary reason those warped
belts are employed is to detect gamma-ray signals as accurately as possible
if they are emitted from the Galactic plane with greater intensity than from other
sky regions.  In addition, the zenith angle
distribution, and hence the primary energies of detected air showers are quite
similar in each warped belt.  This is very important doing the
estimation of the on-plane background intensity at the same energy for both
the on-plane and many off-plane belts, because the gamma-ray energy spectrum
is still unknown. The final reason is that the background estimate, especially 
for the OG, is little affected by the IG plane, because the warped belts 
crossing the IG plane do so diagonally, over a narrow band of longitudes.

\placefigure{fig3}

Figure 4 shows the distribution of the number of events in 4$^\circ$ bin
warped belts. The abscissa represents the right ascension of each warped
belt at the declination 30$^\circ$, which is the same as the latitude of the
Yangbajing site.  The solid lines are the curves fitted to the experimental
data, ignoring the on-plane data.  In this figure, the cases for (a) IG and
(b) OG in the Tibet-III data at 3 TeV and (c) IG and (d) OG in the Tibet-II
data at 10 TeV are shown in the 4$^\circ$ bin analysis.  The anisotropy of
cosmic-ray intensity seen in this figure, with an amplitude of less than
$\pm$1\%, is mainly due to some seasonal long suspensions of operation for
construction and system calibration of the array, and partly due to weather
conditions and a slight, 1$^\circ$ or less, inclination of the site at
Yangbajing.

\placefigure{fig4}

An excess of the on-plane data above the fitted curve would be considered
to be a gamma-ray signal. The signal strength is measured by a standard deviation
of the number of showers, in two or five 2$^\circ$ bin belts, by the formula
$(E-B)/\sqrt{B}$, where $E$ is the number of on-plane events and $B$ is the
estimated number of background events in the on-plane region.  $B$ is
estimated from the solid curve which is obtained by the fitting of
many 2$^\circ$ bin off-plane belts, ignoring the central 16$^\circ$ or
14$^\circ$ widths corresponding to the 4$^\circ$ bin or 10$^\circ$ bin belt
analyses. We intend to deduce the number of on-plane background events as
accurately as possible from this fitting curve of the off-plane data.  The
number of events shown by the solid curve in each subfigure of Fig. 4 is
assumed to be due to galactic cosmic rays.  In this method, an isotropic
extragalactic diffuse gamma-ray component can not be distinguished from the
galactic cosmic rays. We assume that the intensity of the isotropic component is
negligible in comparison with the diffuse gamma rays from the Galactic plane.

Figure 5 shows the deviation distributions of the number of events from the
fitted curves
(the solid ones in Fig. 4) in 2$^\circ$ bin off-plane belts for (a) IG and
(b) OG in the Tibet-III data at 3 TeV and (c) IG and (d) OG in the Tibet-II
data at 10 TeV, respectively, in the case of the 4$^\circ$ bin analysis.
Here, the abscissa represents a significance in the 2$^\circ$ bin off-plane
belts.  The solid curves are best-fit Gaussians with standard deviations of
0.998, 1.007, 0.999 and 1.000, respectively, and where data on the
12$^\circ$ width centered at the Galactic plane are excluded.  In the case
of the 10$^\circ$ bin analysis, almost the same distributions are obtained;
their standard deviations are 1.018, 1.007, 1.013 and
1.000, respectively.  Because the fluctuation of the number of events around
the fitted curve in the off-plane belts is quite natural, i.e. each standard
deviation is almost equal in unity, the solid curves are considered to be
satisfactorily fitted to the experimental data. Thus, we can accurately
estimate the number of on-plane background events.

\placefigure{fig5}

The diffuse gamma-ray intensity from the OG plane is likely to be much lower than
that from the IG plane. The well known gamma-ray point source, the Crab
Nebula, is located at $\alpha=83.\hspace{-1.3mm}^\circ38^\prime,
\delta=22.\hspace{-1.3mm}^\circ01^\prime$ ($l=184.56^\circ,  b=-5.78^\circ$)
with a real angle distance of 5.78$^\circ$ from the central sheet of the OG
plane.  The Crab is in the off-plane belt adjacent to the on-plane belt in
the 10$^\circ$ bin analysis.  In the 4$^\circ$ bin analysis, the Crab is located
between the on-plane and off-plane regions but nearer to the latter.  For that
reason, we omit the two 2$^\circ$ bin warped belts from the off-plane region.
As a result, the data in the 8 warped belts, $-10^\circ < b < 6^\circ$, are
excluded from the background estimation in the 4$^\circ$ bin analysis, while
the 7 warped belts, $-9^\circ < b < 5^\circ$, are excluded in the 10$^\circ$
bin analysis.  Thus, the real angle distance of the Crab is 4.21$^\circ$
from the on-plane region
and 2.22$^\circ$ from the newly defined off-plane region in the 4$^\circ$ bin
analysis, and 1.82$^\circ$ and 1.40$^\circ$ in the 10$^\circ$ bin analysis.
Therefore, the Crab is located sufficiently far from both the on-plane and
off-plane regions, because the angular resolutions of the Tibet-III at 3 TeV
and the Tibet-II at 10 TeV are both about 0.9$^\circ$.

In order to see the adequacy of the fitting and on-plane excess in more
detail, smaller ranges of the right ascension band of about 50$^\circ$
around the Galactic plane are shown in Fig. 6, where subfigures (a), (b),
(c) and (d) correspond to those in Fig. 4.  In each subfigure, the
on-plane and off-plane data are indicated by filled circles and filled
triangles, respectively, and the blank squares indicate the data points excluded
from the background estimation.  Error bars are statistical only.  It
is noted that the error bars are short in the subfigure (b) because the
ordinate scale is compressed by about half compared to the other subfigures.

\placefigure{fig6}

\section{RESULTS and DISCUSSION}

The significance of an on-plane excess, $(E-B)/\sqrt{B}$ measured in
standard deviation of the number of on-plane events, is calculated for
each distribution shown in Fig. 6 and the results are summarized in Table 2.
In Table 2, the results thus obtained are given for the regions of IG
($20^\circ \leq l \leq 55^\circ$) and OG ($140^\circ \leq l \leq
225^\circ$). Mode energies 3 TeV and 10 TeV indicate the analyses of the
Tibet-III and Tibet-II data, respectively. As given in this table, no
significant excess is found, although an excess of +2.52 $\sigma$ is marginal
for IG at 3 TeV in the 4$^\circ$ bin analysis of the Tibet-III data.  We
calculate the upper limits for the gamma-ray intensity using the methods
given by Helene (1983) and Protheroe (1984), specifically for a small excess
or deficit on the Galactic plane.  In this table, $J_\gamma/J_{\rm CR}$
means the flux ratio at 1 $\sigma$ excess, which is identical to
$1/\sqrt{B}$, of the diffuse gamma rays
versus the galactic cosmic rays in the energy region above 3 TeV and above 10
TeV.  A minor component of the isotropic diffuse gamma rays is included
in the galactic cosmic rays because separating them is impossible
in the Tibet air shower array due to the lack of any equipment to reduce the
background hadron initiated air showers.

\vspace{-3mm}

\begin{center}
TABLE 2\\
LIMITS TO DIFFUSE GAMMA RAYS
\end{center}

\begin{tabular}{c|c c c c c c}\hline \hline\\[-5mm]
Inner or Outer&        & Mode  & Signifi- &$\frac{J_\gamma(>E)}{J_{\rm
CR}(>E)}$ {\small at 1$\sigma$}&\multicolumn{2}{c}{$E^2
\frac{dJ_\gamma(>E)}{dE}$}\hspace{0.1mm}{\scriptsize
(cm$^{-2}$s$^{-1}$sr$^{-1}$MeV)}\\
Galactic Plane& Region & Energy&cance&({$\small \equiv 1/\sqrt{B}$}) &
\hspace*{3.5mm}90\% CL & 99\% CL\\
(Region
of\hspace{2mm}$l$)&of\hspace{2mm}$b$&(TeV)&($\sigma$)&(10$^{-4}$)&\hspace*{3.5mm}(10$^{-3}$)&(10$^{-3}$)\\\hline
         &        & 3 &  +2.52&1.95 & 7.6 & 9.6\\[-3mm]
         &$|b|<2^\circ$ & & && &\\[-3mm]
I G      &        & 10 &+1.71&2.43 & 3.0 & 4.0\\[1mm]
{\small (20$^\circ\leq l\leq 55^\circ$)}& & 3 &+1.88&1.23 & 4.0 &
5.3\\[-3mm]
         &$|b|<5^\circ$& & &&& \\[-3mm]
         &        & 10 &+0.81&1.54 & 1.4 & 2.0\\\hline
         &        & 3  &+0.25&1.16 & 2.1 & 3.3\\[-3mm]
         &$|b|<2^\circ$ & & &&&\\[-3mm]
O G      &       & 10 &$-0.63$&1.45 & 0.78 & 1.3\\[1mm]
{\small (140$^\circ\leq l\leq 225^\circ$)}&& 3 &+1.78&0.737& 2.3 &
3.1\\[-3mm]
         &$|b|<5^\circ$& & &&&\\[-3mm]
         &        & 10 &$-0.66$&0.936 & 0.50 & 0.83 \\\hline
\end{tabular}

\vspace*{3mm}

\par
In Table 2, the intensity upper
limits are also given as 90\% and 99\% confidence level (CL),
calculated from the above flux ratio and assuming a
differential spectral index of 2.4 for the diffuse gamma rays, and utlizing
the all-particle energy spectrum of the galactic cosmic rays
recently compiled by Apanasenko et al. (2001).

\par
Figure 7 shows the 99\% CL upper limits thus obtained for diffuse gamma rays
from the inner Galactic plane, $20^\circ \leq l \leq 55^\circ$ and $|b| \leq
2^\circ$, at energies around 3 TeV (T3 for Tibet-III) and 10 TeV (T2 for
Tibet-II). In this figure the EGRET data \cite{hunt97}, $315^\circ \leq l
\leq 45^\circ$ and $|b| \leq 2^\circ$, are plotted.  The Cherenkov data are
also plotted, including
Whipple's 99.9\% CL upper limit above 500 GeV \cite{lebo00} at the region of 
$38.5^\circ \leq l \leq 41.5^\circ$ and $|b| \leq 2^\circ$,
and HEGRA-IACT's 99\% CL upper limit above 1 TeV \cite{aha01} in a similar
region of $38^\circ \leq l \leq 43^\circ$ and $|b| \leq 2^\circ$.
The theoretical curve calculated by Berezinsky et al. (1993) for
$\pi^\circ\rightarrow2\gamma$ due to the collision of cosmic-ray hadrons
with ISM is drawn by a solid curve (BGHS) which is for the region of
$20^\circ \leq l \leq 55^\circ$ and $|b| \leq 2^\circ$, deduced from their
original paper which
is based on the matter density distribution compiled by Fichtel \& Kniffen
(1984) and Bloemen et al. (1984).
\par
For inverse Compton gamma rays induced by energetic electrons, the
calculation by Porter \& Protheroe (1996) is shown by dashed curves for
source electron spectral indices of 2.0 (PP2.0) and 2.4 (PP2.4) in the
direction $l=0^\circ$ and $b=0^\circ$, the Galactic center.  Similar
theoretical curves calculated by Tateyama \& Nishimura (2001) are shown by
dot-dashed lines with source spectral indices of 2.0 (TN2.0) and 2.4 (TN2.4)
in the direction $l= 0^\circ$ and $|b| \leq 2^\circ$. The density
distribution of ISPF in their calculations is based on the compilations by
Bloemen (1985) and Mathis et al. (1983).  The curves from Tateyama \&
Nishimura (2001) are consistent with the estimations by Porter \& Protheroe
(1996) considering the different region of $|b|$.  The present Tibet data,
especially at 10 TeV, together with the
HEGRA-IACT data, give the most stringent upper limit for the IC model, although
these data can not clearly rule
out the IC model with a source electron spectral index of 2.0.

\placefigure{fig7}

Figure 8 shows the present results as 99\% CL upper limits (upper bars of
T3 and T2) for diffuse gamma rays from the outer part of the Galactic
plane, $140^\circ \leq l \leq 225^\circ$ and $|b| \leq 2^\circ$, for the
energy ranges around 3 TeV (T3 for Tibet-III) and 10 TeV (T2 for Tibet-II).
In this figure, the 90\% CL upper limits (lower bars) are
compared with the CASA-MIA 90\% CL upper limits, which are based upon muon-poor
air showers \cite{bori98} at about 140 TeV-1.3 PeV, from the OG plane of
$50^\circ \leq l \leq
200^\circ$ and $|b| \leq 2^\circ$.  The CASA-MIA data can rule out the IC
model with index 2.0 without acceleration energy cutoff, but can not rule out
the case with an energy cutoff at 100 TeV.  In this figure, the theoretical
curve by Berezinsky et al. (1993) is also shown for $\pi^\circ \rightarrow
2\gamma$ component (BGHS) in the region of $140^\circ \leq l \leq 225^\circ$
and $|b| \leq 2^\circ$.
The IC gamma rays calculated by Porter \&
Protheroe (1996) for the region $50^\circ \leq l \leq 200^\circ$ and $|b|
\leq 10^\circ$ for source spectral indices of 2.0 (PP2.0) and 2.4 (PP2.4) are
shown, as well as the ones by Tateyama \& Nishimura (2001) in the region
of $l=180^\circ$ and $|b| \leq 2^\circ$ for source spectral indices of 2.0
(TN2.0) and 2.4 (TN2.4). The present data are also not sufficient to rule
out the IC model with a spectral index of 2.0.

\placefigure{fig8}

Next we discuss some considerations regarding
our method of data analysis and its results.  First, a difference of the
average shower size between gamma-ray and proton initiated showers with the
same energy at the Yangbajing site (606 gcm$^{-2}$) will produce different
initial proton and gamma-ray energies for the same observed shower size.
By employing the
subroutine package GENAS of Kasahara \& Torii (1991), the average
incident gamma-ray
energy is roughly estimated to be lower by 18\% at 3 TeV and lower by 23\%
at 10 TeV than that for protons of the same shower size, taking
the median zenith angle 24.1$^\circ$ (atmospheric depth 664 gcm$^{-2}$) into
account for generally observed shower events.  The median zenith angles
are 27.2$^\circ$ and 22.8$^\circ$ (681 gcm$^{-2}$ and 657 gcm$^{-2}$) for
the showers arriving from the inner and outer Galactic plane, respectively.
Thus, the energy decline rate in gamma-initiated showers is somewhat reduced
for the IG plane and magnified a little for the OG plane.  On the other hand,
the spectral index of the diffuse gamma rays is probably smaller than the 2.7 of
the galactic cosmic rays at energies around 10 TeV.  This produces an opposite
offset of the observed gamma-ray energy. which shifts to a higher value, depending
on the spectral index.  If the gamma-ray spectral index is 2.4, as assumed
in the derivation of the intensity upper limit in Table 2, gamma-ray energy
goes up about 10\% or more.  The effects just described act to offset
each other.  Thus we use the observed energies,
3 TeV for the Tibet-III and 10 TeV for the Tibet-II, for the primary
energies determined for the generally observed air showers at Yangbajing site.

Second, the sky regions searched for Galactic diffuse gamma rays have a shape
like a convex lens along the Galactic plane as shown in Fig. 2.  Such a lens
shaped region is inevitable in our method, employing the warped belts along
the Galactic plane in equatorial coordinates.  The maximum thickness of
the lens is 3.6$^\circ$ or 8.9$^\circ$, but the simple mean of the range used
is 3.5$^\circ$ or 8.7$^\circ$ for IG and 2.7$^\circ$ or 6.8$^\circ$ for OG in
the 4$^\circ$ bin and 10$^\circ$ bin analyses, respectively.  Taking the
number density of events into account, the weighted mean thickness
becomes 3.5$^\circ$ or 8.7$^\circ$ for IG and 2.9$^\circ$ or 7.2$^\circ$ for
OG, respectively.  As already described, the data in the two 2$^\circ$ bin
belts has been excluded from the off-plane region for OG in order to minimize
the influence of the Crab Nebula.  Other strong gamma-ray sources,
Geminga ($l=195.03^\circ, b=4.83^\circ$) and IC443 ($l=188.83^\circ,
b=3.07^\circ$) are involved
in the on-plane region of the OG plane in the 10$^\circ$ bin analysis. No
TeV gamma-ray signal has, however, been obtained from these candidate sources
by any surface air shower arrays or Cherenkov telescopes. Therefore, no
correction for the influence of these candidates is necessary.

Third, according to the EGRET data \cite{hunt97} with $|b| \leq 2^\circ$,
gamma-ray intensity shows a decline in the galactic longitude region from
$l=25^\circ$ to $65^\circ$ in IG for every energy ranges, 30-100 MeV, 100-300 MeV, 
300-1000 MeV and above 1000 MeV.  This tendency is also seen in
SAS-2 and COS-B data (see Bertsch et al. 1993).  The intensity shows a rather
flat top in the central range, $330^\circ \leq l \leq 25^\circ$.  In our analysis,
the inner Galactic region is located just at this region of declining intensity.
The Whipple \cite{lebo00} and HEGRA \cite{aha01} data
also lie partially in this region, at around $l=40^\circ$.  If the intensity
decline around 1 GeV is caused by the density distribution of ISPF, it is
expected that the gamma-ray intensity shows similar behavior
in the TeV region.  We should compare the experimental data with the
reduced intensity of about 80 \% of the theoretical curves, which have
been calculated for the central flat top region.  This is the reason that the
experimental data are not sufficient to rule out the inverse Compton model
with a source spectral index of 2.0.

Theoretical calculations give an indication that, if the source electron
spectral index in the 10 GeV to 10 TeV energy region is smaller than 2.4,
the diffuse gamma rays in the TeV region are mainly generated by IC scattering.
In that case, it is essential to fix the diffuse gamma-ray intensity to
determine the source electron spectrum in the Galactic plane.  This can
suggest the strength of shock acceleration, and clarifies the electron
propagation process in the Galactic disc through a comparison with the
direct observation of local electrons, and also gives an estimate of
average magnetic field in the source region by examining the consistency
with the radio synchrotron intensity.

The extension of the Tibet-III array is expected to be completed by the end of
2002; its effective inner area will become about 1.5 times larger than at present,
and 1.15 times larger than the Tibet-II inner area.  If the new
Tibet-III array continues for running three years without long suspension,
statistics will increase to 4-5 times the present data at both 3 TeV
and 10 TeV.  The resulting statistical reduction in the upper limits, by a
factor of 2 or more, will be more closely comparable with theoretical
models, and can also give a significant upper limit at even higher energies,
e.g., at 20 TeV, where the upper limit is relatively more sensitive to the
acceleration energy cutoff in the IC model.

\acknowledgments
  This work is supported in part by Grants-in-Aid for Scientific
Research on Priority Areas from the Ministry
of Education, Culture, Sports, Science and Technology in Japan and
from the Committee of the Natural Science Foundation and the Academy of
Sciences in China.


\clearpage
\begin{figure}
\epsscale{.7}
\plotone{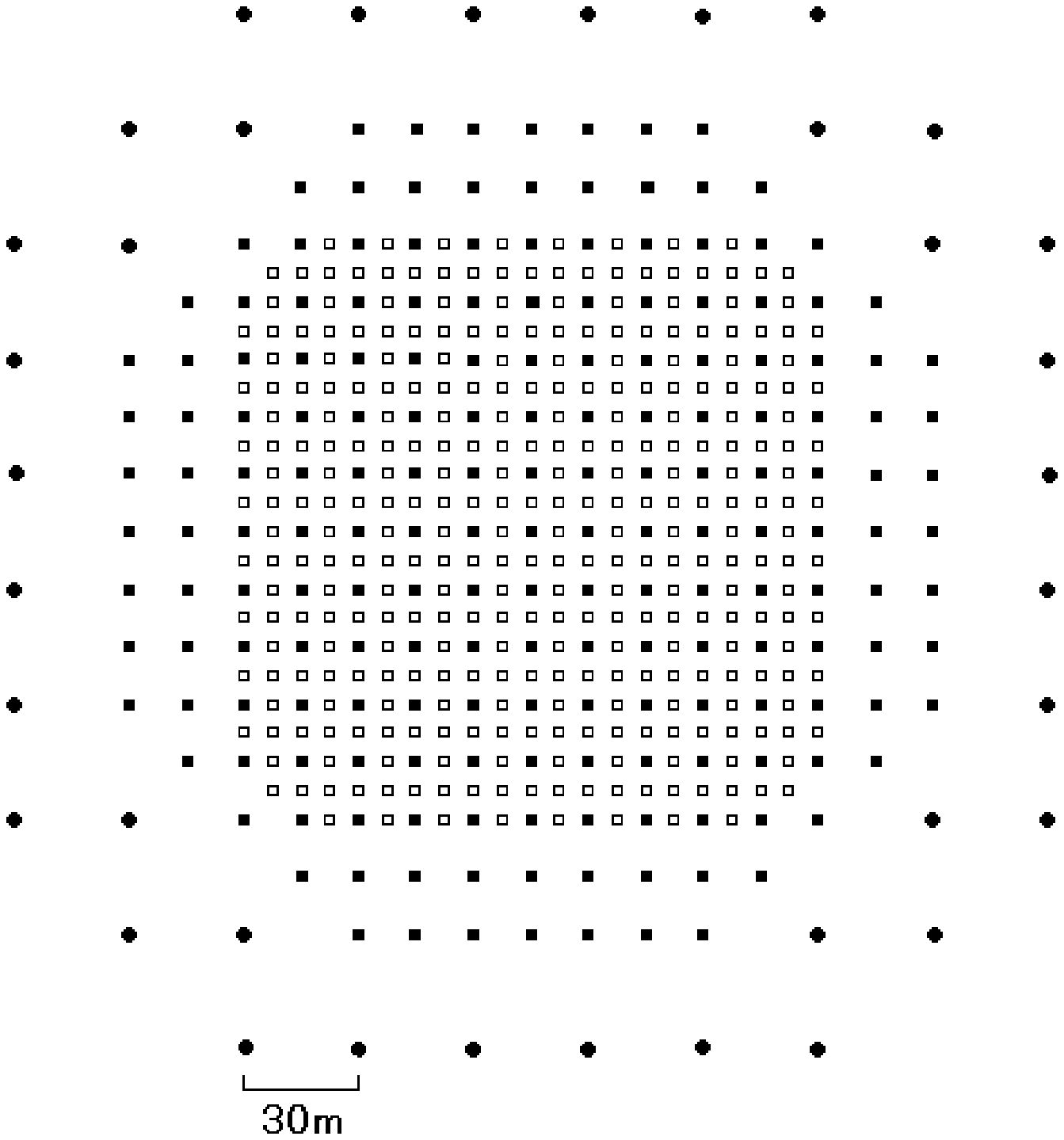}
\caption{Array map of the Tibet-II and Tibet-III at Yangbajing.
Filled squares are detectors in the Tibet-II (1995-1999) and open ones
are added detectors to construct the Tibet-III (1999-2002).  Filled
circles are density detectors equipped with a wide dynamic range phototube.
\label{f1}}
\end{figure}

\begin{figure}
\plotone{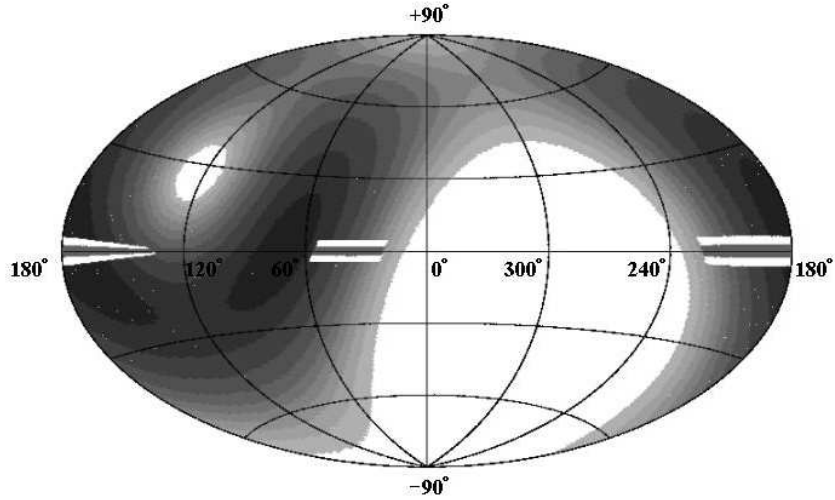}
\caption{Tibet-III exposure map in the galactic coordinates for showers
with zenith angles $\theta \leq 50^\circ$.  The event density increases from
light gray to dark gray.  Boundaries of searched sky region for diffuse
gamma rays from the Galactic plane are indicated by narrow gray ($|b| \leq
2^\circ$) and white plus gray ($|b| \leq 5^\circ$) along the Galactic plane.
\label{f2}}
\end{figure}

\begin{figure}
\plotone{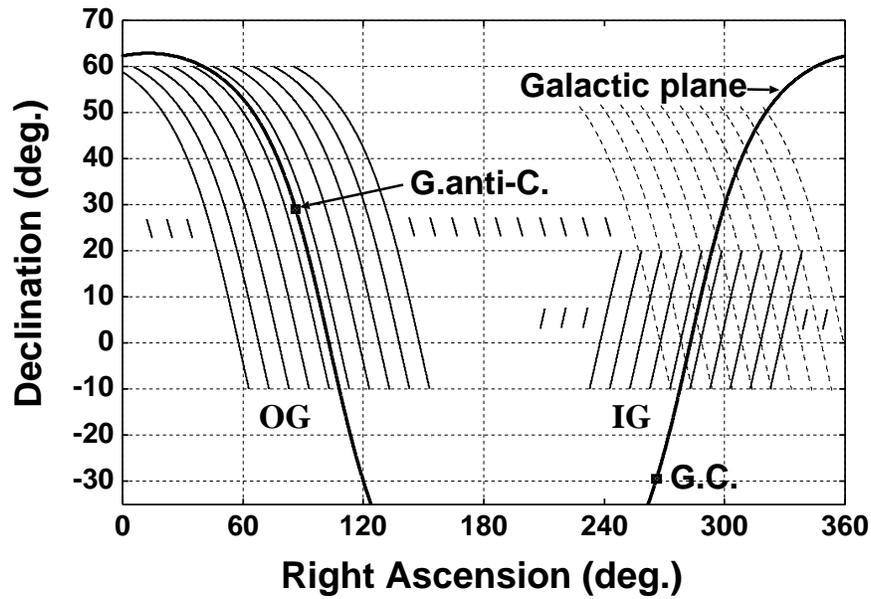}
\caption{Warped belts of 10$^\circ$ width in the right ascension along
the Galactic plane for IG near the Galactic center and OG including the
Galactic anticenter. \label{f3}}
\end{figure}

\begin{figure}
\plotone{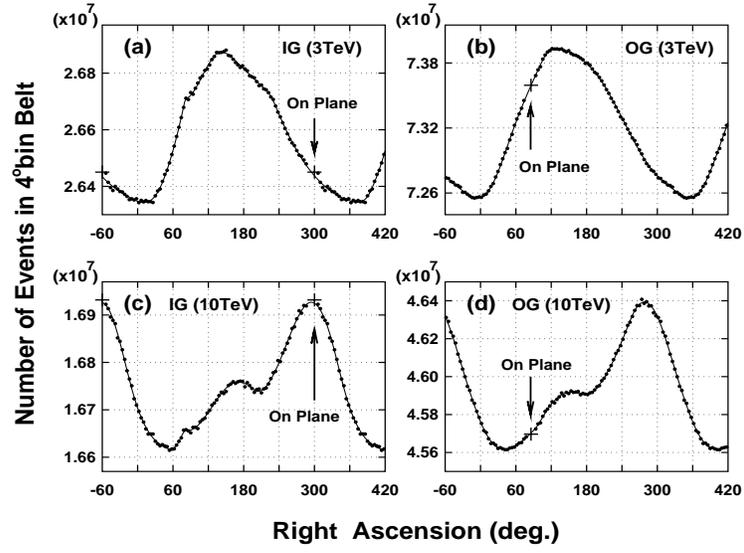}
\caption{Distributions of number of events in 4$^\circ$ bin warped belts
for (a) IG and (b) OG in Tibet-III data at 3 TeV and (c) IG and (d) OG in
Tibet-II data at 10 TeV.   Abscissa represents the right ascension of each
belt at the declination 30$^\circ$.  Solid lines are curves fitted to the
experimental data.  In each subfigure, the on-plane data point is shown by a
large + for the 4$^\circ$ bin analysis. \label{f4}}
\end{figure}

\begin{figure}
\plotone{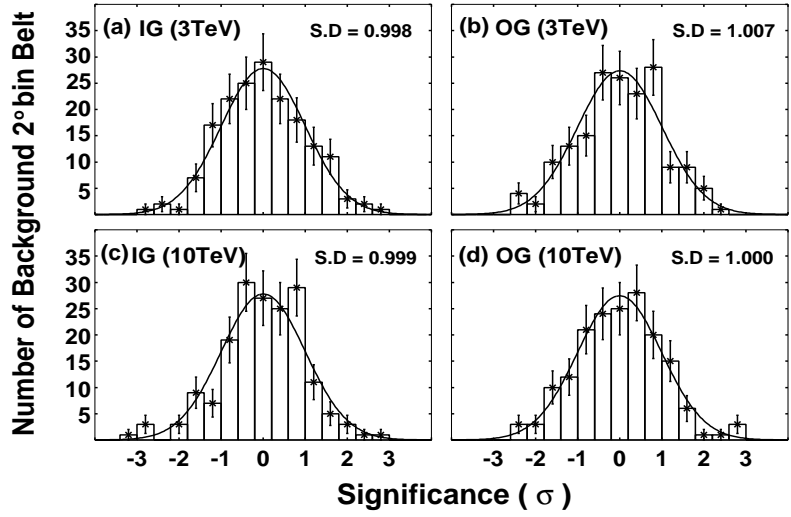}
\figcaption{Deviation distributions of 2$^\circ$ bin off-plane data from the
fitted curves for (a) IG and (b) OG for the Tibet-III data at 3 TeV, and (c)
IG and (d) OG for the Tibet-II data at 10 TeV in the 4$^\circ$ bin analysis.
For IG, the data in the central 12$^\circ$ are omitted from the off-plane
data.  For OG, the data in the range $-10^\circ \leq l \leq 6^\circ$ are
also excluded to minimize the Crab's influence in the 4$^\circ$ bin
analysis.  Solid curves are best-fit Gaussians with standard deviation
designated in each subfigure. \label{f5}}
\end{figure}

\begin{figure}
\plotone{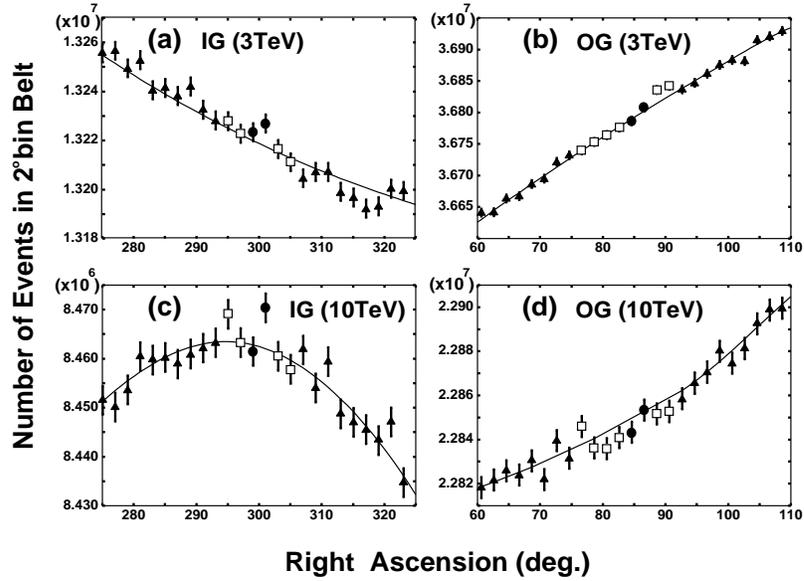}
\caption{Distributions of the number of events in the right ascension
range of 50$^\circ$.  Subfigures (a), (b), (c) and (d) correspond to
those in Fig. 4.  Abscissas are the same in Fig. 4.  On-plane data are shown
by filled circles, off-plane data by filled triangles, and open squares
indicate data omitted from the background estimation, as described in the text.
The fitted solid curves in each subfigure are used for estimation of the
number of on-plane background events $B$.  $E-B$ is the excess number of
events over the solid curves in the on-plane belts. \label{f6}}
\end{figure}

\begin{figure}
\plotone{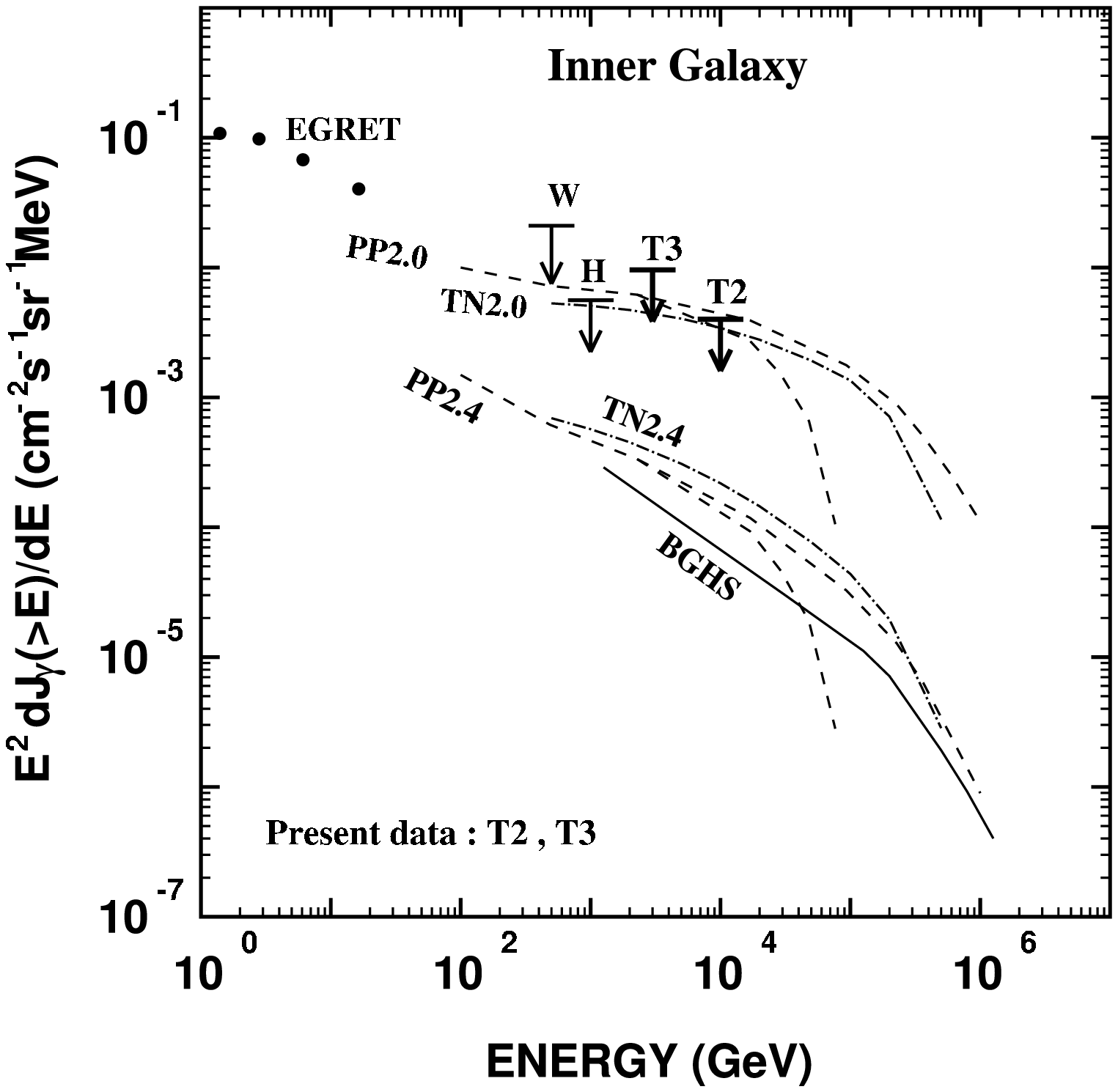}
\figcaption{Diffuse gamma rays from the inner Galaxy (IG).  Present data are
labeled by T2 and T3 as 99\% CL upper limits, assuming a gamma-ray spectral
index 2.4.  W and H indicate the Whipple's 99.9\% CL \cite{lebo00} and
HEGRA's 99\% CL \cite{aha01} upper limits with IACT.  Theoretical curves are
labeled by initials of the authors names. BGHS represents
$\pi^\circ\rightarrow2\gamma$ by Berezinsky et al. (1993).  PP and TN are
given by Porter \& Protheroe (1996) and Tateyama \& Nishimura (2001) for the
inverse Compton. The numerals 2.0 or 2.4 following PP and TN indicate
the electron source differential spectral indices. \label{f7}}
\end{figure}

\begin{figure}
\plotone{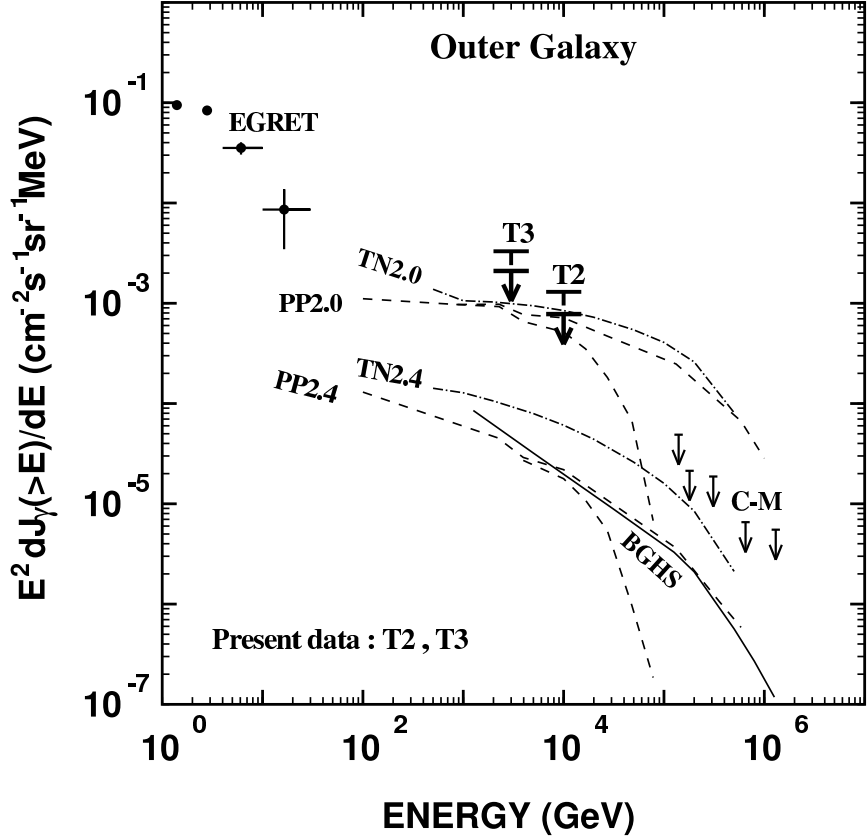}
\figcaption{Diffuse gamma rays from the outer Galaxy (OG).  Present data are
labeled by T2 and T3 as 99\% CL (upper bars) and 90\% CL (lower bars) upper
limits, assuming the same spectral index 2.4.   The latter are compared with
CASA-MIA 90\% CL upper limits (labeled by C-M), based on muon-poor
air shower data \cite{bori98}.  Theoretical curves and their labels are the
same as in Fig. 7. \label{f8}}
\end{figure}
\end{document}